\documentclass[onecolumn,showpacs,preprintnumbers,amsmath,amssymb,superscriptaddress]{revtex4}
\usepackage{graphics,psfrag}
\usepackage{rotating}
\usepackage[usenames]{color}
\usepackage{amsmath}
\usepackage{bbm}
\usepackage{ulem} 
\newcommand{\be}{\begin{eqnarray}}
\newcommand{\ee}{\end{eqnarray}}

\newcommand{\non}{\nonumber\\}

\newcommand{\del}{\partial}

\newcommand{\C}{{\cal C}}

\newcommand{\pr}{^\prime}

\newcommand{\Gm}{\Gamma}
\newcommand{\rw}{\rightarrow}

\newcommand{\bk}{\bar{K}}
\newcommand{\eps}{\epsilon}

\newcommand{\Sgs}{\Sigma^*}
\newcommand{\mev}{\textrm{ MeV}}
\newcommand{\Ls}{{\Lambda^*(1520)}}
\newcommand{\Lss}{$\Lambda^*(1520)\;$}

\begin{document}
\title{Radiative decay of the $\Lambda^*(1520)$}

\author{M. \surname{D\"oring}}
\email{doering@ific.uv.es}
\affiliation{Departamento de F\'{\i}sica Te\'orica and IFIC,
Centro Mixto Universidad de Valencia-CSIC,\\
Institutos de
Investigaci\'on de Paterna, Aptd. 22085, 46071 Valencia, Spain}
\author{E. \surname{Oset}}
\email{oset@ific.uv.es}
\affiliation{Departamento de F\'{\i}sica Te\'orica and IFIC,
Centro Mixto Universidad de Valencia-CSIC,\\
Institutos de
Investigaci\'on de Paterna, Aptd. 22085, 46071 Valencia, Spain}
\author{Sourav \surname{Sarkar}}
\email{sourav@veccal.ernet.in}
\affiliation{Departamento de F\'{\i}sica Te\'orica and IFIC,
Centro Mixto Universidad de Valencia-CSIC,\\
Institutos de
Investigaci\'on de Paterna, Aptd. 22085, 46071 Valencia, Spain}
\affiliation{Variable Energy Cyclotron Centre, 1/AF Bidhannagar, Kolkata-700064, India}

\begin{abstract}
A recently developed non-perturbative chiral approach to dynamically generate the $3/2^-$ baryon resonances
has been extended to investigate the radiative decays $\Lambda^*(1520)\to\gamma\Lambda(1116)$ and $\Lambda^*(1520)\to\gamma\Sigma^0(1193)$. We show that the \Lss decay into $\gamma\Lambda$ is an ideal test for the need of extra components of the resonance beyond those provided by the chiral approach since the largest meson-baryon components give no contribution to this decay. The case is different for $\gamma\Sigma$ decay where the theory agrees with experiment, though the large uncertainties of these data call for more precise measurements. 
\end{abstract}
\maketitle
\section{Introduction}
New light has been brought in the study of the meson-baryon interaction by the unitary extensions of chiral perturbation theory $U\chi PT$, showing that some well-known resonances qualify as being dynamically generated. In this picture the Bethe-Salpeter resummation of elementary interactions derived from chiral Lagrangians guarantees unitarity and leads at the same time to genuine non-perturbative phenomena such as poles of the scattering amplitude in the complex plane which can be identified with resonances. Coupled channels play an essential role in this scheme, as the chiral Lagrangians provide the corresponding transitions within the multiplets, and even physically closed channels can contribute effectively. 
It is interesting to note that, even without chiral Lagrangians, the use of basic interactions for the coupled channels calls for an interpretation of some resonances like the $\Lambda(1405)$ as quasibound states  of the scattering problem \cite{dalitz,Jennings:1986yg}.

After earlier studies in this direction explaining the $\Lambda(1405)$ and the $N^*(1535)$ as meson-baryon $(MB)$ quasibound states \cite{Kaiser:1995cy,Kaiser:1996js,kaon,Nacher:1999vg,Oller:2000fj} from the interaction of the meson octet of the $\pi$ with the baryon octet of the $N$, new efforts have been undertaken \cite{Kolomeitsev:2003kt, Sarkar:2004jh} to investigate the low lying $3/2^-$ baryonic resonances which decay in $s$-wave into $0^-$ mesons $(M)$ and $3/2^+$ baryons $(B^*)$ of the decuplet. The latter particles, the $0^-$ mesons and $3/2^+$ baryons, provide the building blocks of the coupled channels 
needed in the study of the meson-baryon $s$-wave interaction in the $3/2^-$ channel. A parameter free Lagrangian accounts for this interaction at lowest order and the model exhibits poles in the different isospin and strangeness channels in the complex $\sqrt{s}$-plane, which have been identified with resonances such as $\Lambda^*(1520)$, $\Sigma^*(1670)$, $\Delta^*(1700)$, etc.

However, the $3/2^-$ resonances have also large branching ratios for $(0^-, \;1/2^+)$ MB decays in $d$-wave, in many cases being even larger than the $s$-wave branching ratio due to larger available phase space. For a realistic model that can serve to make reliable predictions in hadronic calculations, the $d$-wave channels corresponding to these decays should be included as has been been done recently in Ref. \cite{Sarkar:2005ap} for one of the $3/2^-$ resonances from Ref. \cite{Sarkar:2004jh}, the $\Lambda^*(1520)$. For the $MB\to MB^*$ $s$-wave to $d$-wave and $MB\to MB$ $d$-wave to $d$-wave transitions, chiral symmetry does not fix the coupling strength so that free parameters necessarily enter the model. On the other hand, this freedom allows for a good reproduction of $d$-wave experimental data for $\overline{K}N\to\overline{K}N$ and $\overline{K}N\to\pi\Sigma$ via the $\Lambda^*(1520)$, see Ref. \cite{Sarkar:2005ap,new_Oset_Sarkar_Roca}. Once the free parameters are determined by fitting to the experimental data of these reactions, the predictivity of the model can be tested for different data sets as has been done in Ref. \cite{new_Oset_Sarkar_Roca} for the reactions $K^-p\to\pi^0\pi^0\Lambda$, $K^-p\to\pi^+\pi^-\Lambda$, $\gamma p\to K^+K^- p$, and $\pi^-p\to K^0K^-p$, finding in all cases good agreement with data.

In the present study we extend the chiral coupled channel approach --- without introducing new parameters --- to investigate the radiative decays $\Lambda^*(1520)\to\gamma\Lambda(1116)$ and $\Lambda^*(1520)\to\gamma\Sigma^0(1193)$ for which new experimental results exist \cite{Taylor:2005zw}. These reactions are of particular interest because they provide further insight into the nature of the $\Lambda^*(1520)$: 
A pure dynamically generated resonance would be made out of meson-baryon components, a genuine resonance would be made of three constituent quarks, but an admixture of the two types is possible and in the real world non-exotic resonances have both components, although, by definition, the meson-baryon components would largely dominate in what we call dynamically generated resonances. Yet, even in this case it is interesting to see if some experiments show that extra components beyond the meson-baryon ones are called for.

The radiative decay of the $\Lambda^*(1520)$ provides a clear example of this: in one of the decays,  $\Lambda^*(1520)\to\gamma\Lambda(1116)$, isospin symmetry filters out the dominant channels $\pi\Sigma^*$ and $\pi\Sigma$ of the present approach so that a sizable fraction of the partial decay width could come from a genuine three quark admixture. In contrast, these dominant channels add up in the isospin combination for the $\Lambda^*(1520)\to\gamma\Sigma^0(1193)$ reaction, and a match to the experimental data would point out the dominant component for this channel being the quasibound meson-baryon system in coupled channels. 

This situation is opposite to the quark model picture of Ref.  \cite{Kaxiras:1985zv} where the decay into $\gamma\Sigma^0(1193)$ is suppressed. This appears as a consequence of selection rules occurring in the limit in which only strange quarks are excited to $p$-wave bag orbits. Indeed, the photon de-excitation of the strange quark with a one-body operator does not affect the isospin of the $u$, $d$ quarks and hence $I=1$ baryons in the final state are forbidden in this limit \cite{Kaxiras:1985zv}.  However, as said above, it is precisely  the $\gamma\Sigma^0(1193)$ final state which in our hadronic interaction picture appears enhanced. We should also mention other quark models \cite{Darewych:1983yw,Warns:1990xi,Umino:1991dk,Umino:1992hi} that enlarge and complement Ref. \cite{Kaxiras:1985zv}, as well as algebraic models \cite{Bijker:2000gq} where the $\Lambda(1520)$ radiative decay has been evaluated.

In the quark model of Ref. \cite{Kaxiras:1985zv} it is shown that the partial decay widths of the $\Lambda^*$ depend sensitively on the $q^4\overline{q}$ admixture which would correspond to meson-baryon components and, thus, could be related to the dynamically generated $\Lambda^*$.

\section{Formulation}  
Before we proceed further, and in order to justify the procedure we follow, we present a general perspective of the ideas and techniques employed in the approach.

The first remark is that the method of dynamically generating resonances is not a tool to describe all resonances of the particle data group (PDG) \cite{Eidelman:2004wy}. Restricting ourselves to the baryonic resonances, thus far, only the low lying $1/2^-$ and $3/2^-$ resonances qualify as such. The quantum numbers of these resonances are such that they can also be in principle interpreted as ordinary three constituent quark states with one quark in a $p$-wave which means that one should be ready to accept some three constituent quark components in the wave function. Conversely, the coupling of meson-baryon components to a seed of three constituent quarks is also unavoidable, as given for instance from the existence of meson-baryon decay channels. Nature will make this meson-baryon cloud more important in some cases than others, and those where the dress of meson cloud overcomes the original three constituent quark seed are candidates to be well described in the chiral unitary approach and appear as what we call dynamically generated resonances where the three constituent quark components are implicitly assumed to be negligible.

Then the question arises, which are the mesons and baryons that are used as building blocks in the chiral unitary approach and which can be dynamically generated. The answer to this is provided by exploiting the chiral theories in the large $N_c$ limit. The dynamically generated resonances appear as a solution of the Bethe-Salpeter equation and hence it is 	the iteration of the kernel through loop diagrams that will lead to the appearance of these resonances. But these are sub-leading terms in the large $N_c$ counting that vanish in the limit of $N_c\to\infty$. Hence, the dynamically generated resonances disappear in a theoretical scheme when $N_c\to\infty$ and the resonances that remain are what we call genuine ones. In this sense, the $\Delta(1232)$ (and other baryons of the decuplet) is a genuine resonance which appears degenerate with the nucleon in the large $N_c$ limit \cite{Jenkins:1991es}. This statement might seem to clash with a well-known historical fact, the dynamical generation of the $\Delta(1232)$ from the iteration of the crossed nucleon pole term in the Chew and Low theory \cite{Chew:1955zz}. However, attractive as the idea has always been, the input used in this approach, in particular the simplified $\pi NN$ coupling, is at odds with present chiral Lagrangians and hence that old idea is no longer supported in present chiral approaches. 
A more modern and updated formulation of the problem, according with requirements of chiral dynamics is given in \cite{Meissner:1999vr}. There, the $\Delta$, which qualifies as a genuine resonance, appears through a Castillejo, Dalitz, Dyson pole \cite{Castillejo:1955ed} in the $N/D$ formulation of \cite{nsd}.

A very important work on the meaning of the large $N_c$ limit and the classification of states into dynamically generated or genuine resonances is Ref. \cite{Pelaez:2003dy}, where the author shows what large $N_c$ means in practice, with some subtleties about the strict $N_c=\infty$. At the same time one shows  that the $\rho$ meson qualifies as a genuine resonance while the $\sigma$, $f_0(980)$, and $a_0(980)$ qualify as dynamically generated. 

Next we discuss an issue of relevance which is the relationship of the $N/D$ method and the Bethe-Salpeter equation. This has been discussed in Ref. \cite{nsd} and \cite{Oller:2000fj} but we summarize the problem here for the sake of clarity and completeness.

We start from the equation of unitarity in coupled channels and we shall work in $s$-wave for simplicity (generalization to other partial waves can be seen in \cite{nsd}). Unitarity in coupled channels is written as
\begin{equation}
{\rm Im} T_{i,j} = T_{i,l} \sigma_l T^*_{l,j}
\end{equation}
where $\sigma_i \equiv 2M_l q_i/(8\pi \sqrt{s})$, with $q_i$  the modulus of the 
c.m.
three--momentum, and the subscripts $i$ and $j$ refer to the physical 
channels.
 This equation is most efficiently written in terms of the inverse 
amplitude as
\begin{equation}
\label{uni}
\hbox{Im}~T^{-1}(\sqrt{s})_{ij}=-\sigma(\sqrt{s})_i \delta_{ij}~.
\end{equation}
The unitarity relation in Eq. (\ref{uni}) gives rise to a cut in the
$T$--matrix of partial wave amplitudes which is usually called the 
unitarity or right--hand
cut. Hence, one can write down a dispersion relation for $T^{-1}(\sqrt{s})$
\begin{equation}
T^{-1}(\sqrt{s})_{ij}=-\delta_{ij}\left\{\widetilde{a}_i(s_0)+
\frac{s-s_0}{\pi}\int_{s_{i}}^\infty ds'
\frac{\sigma(s')_i}{(s'-s-i\epsilon)(s'-s_0)}\right\}+{V}^{-1}(\sqrt{s})_{ij} ~,
\label{dispersion}
\end{equation}
where $s_i$ is the value of the $s$ variable at the threshold of channel 
$i$ and
${V}^{-1}(\sqrt{s})_{ij}$ indicates other contributions coming from 
local and
pole terms, as well as crossed channel dynamics but {\it without}
the right-hand cut. These extra terms
can be taken directly from chiral perturbation theory ($\chi PT$)
after requiring the {\it matching} of the general result to the $\chi 
PT$ expressions.
Note also that
\begin{equation}
g(s)_i=\widetilde{a}_i(s_0)+ \frac{s-s_0}{\pi}\int_{s_{i}}^\infty ds'
\frac{\sigma(s')_i}{(s'-s-i\epsilon)(s'-s_0)}
\label{rem2}
\end{equation}
is the familiar scalar loop integral.

One can further simplify the notation by employing a matrix formalism.
Introducing the
matrices $g(s)={\rm diag}~(g(s)_i)$, $T$ and $V$, the latter 
defined in
terms
of the matrix elements $T_{ij}$ and $V_{ij}$, the $T$-matrix 
can be written as:
\begin{equation}
\label{t}
T(\sqrt{s})=\left[I-V(\sqrt{s})\cdot g(s) \right]^{-1}\cdot V(\sqrt{s})
\end{equation}
which can be recast in a more familiar form as
 \begin{equation}
T(\sqrt{s})=V(\sqrt{s})+V(\sqrt{s}) g(s) T(\sqrt{s}).
\label{rem1}
\end{equation}
This equation has the formal appearance of the Bethe-Salpeter equation (BSE) and it is indeed this equation. However, there is a peculiar feature worth noting: the term $VgT$ of the equation is a product of functions $V(\sqrt{s}),\;g(s)$, and $T(\sqrt{s})$ while in the BSE using an ordinary ${\vec r}$ dependent potential, this term has an explicit $d^4 q$ integration involving $V$ and $T$ half off-shell. The appearance of $V$ and $T$ on-shell in Eq. (\ref{rem1}) is a simple consequence of the dispersion relation of Eq. (\ref{dispersion}).

Note that $g(s)$ of Eq. (\ref{rem2}) is nothing but the $d^4 q$ integral of a meson and baryon propagator (the check of the imaginary part is immediate), hence in simple words we can say that the dispersion relation justifies a BSE in which the $V$ and $T$ are factorized on-shell outside the integral of the $VgT$ term. Generalization of this technique to higher partial waves is done in Ref. \cite{nsd}. In this case, there is a subtraction polynomial instead of the subtraction constant of Eq. (\ref{rem2}), but in a narrow region around a resonance this can be taken as a constant. 

There is  a caveat in the argument given above: Eq. (\ref{dispersion}) contains only the contribution of the imaginary part of the amplitude corresponding to the right-hand, physical cut. The unphysical, or left-hand cut contribution is not taken into account. Therefore, there is an approximation involved. Yet, this is an approximation which is kept under control. In \cite{nsd} a test was done of the contribution of the left-hand cut in meson-meson scattering with the conclusion that the contribution is small. But more important: It is weakly energy-dependent in the region of physical energies. This is the key to the success of the method explored here, since any constant contribution in a certain range of energies can be accommodated in terms of the subtraction constant that appears in the $g(s)$ function of Eq. (\ref{rem2}) (see also a detailed discussion of the contribution of the left-hand cut in $\pi N$ scattering in \cite{Meissner:1999vr}). This finding is not unique to the former procedure but in some works  \cite{Gross:1992tj,Surya:1995ur} the crossed nucleon pole terms in $\pi N$ scattering, which would lead to the left-hand cut contribution in the dispersion relation, are approximated by a local term. 

The techniques discussed in this section have been applied successfully to $\overline{K}N$ interaction in $s$-wave \cite{kaon} and $p$-waves \cite{Jido:2002zk}. In this latter work, the kernel, $V$, has contact terms and pole terms corresponding to the $\Lambda$, $\Sigma$, and $\Sigma^*(1385)$ particles. A similar procedure is done in \cite{Oller:2000fj} also for $\overline{K}N$ scattering and in \cite{Meissner:1999vr} in the $\pi N$ scattering case. The quality of the results and the sophistication of that latter model is equivalent to that of other successful relativistic approaches to $\pi N$ like \cite{Gross:1992tj,Surya:1995ur}, and fewer parameters are needed. In the case of the $\overline{K}N$ interaction of \cite{kaon} and \cite{Jido:2002zk} a quite good description of the data was obtained with only one parameter. 

\subsection{$s$-wave channels}
\label{sec:form}
Following Ref. \cite{Sarkar:2004jh}, we briefly recall how the $\Lambda^*(1520)$ appears as a dynamically
generated resonance 
in the $s$-wave interaction of the $3/2^+$ baryon decuplet with the $0^-$ meson octet. 
The lowest order term of the chiral Lagrangian relevant for the interaction is given 
by~\cite{Jenkins:1991es} (we use the metric
$g^{\mu\nu}={\rm diag}(1,-1,-1,-1)$) 
\be
{\cal L}=-i\bar T^\mu {\cal D}\!\!\!\!/ T_\mu 
\label{lag1} 
\ee
where $T^\mu_{abc}$ is the spin decuplet field and $D^{\nu}$ the covariant derivative
given by
\be
{\cal D}^\nu T^\mu_{abc}=\del^\nu T^\mu_{abc}+(\Gm^\nu)^d_aT^\mu_{dbc}
+(\Gm^\nu)^d_bT^\mu_{adc}+(\Gm^\nu)^d_cT^\mu_{abd}
\ee
where $\mu$ is the Lorentz index, $a,b,c$ are the $SU(3)$ indices, and $\Gamma^\nu$ the vector current.
Let us recall the identification of the $SU(3)$ components
of $T$ to the physical states~\cite{Butler:1992pn,lutz3}:
\be
&&T^\mu=T_{ade} u^\mu,\quad \overline{T}_\mu=\overline{T}^{ade}\overline{u}_\mu,\non
&&T_{111}=\Delta^{++},\;T_{112}=\frac{\Delta^+}{\sqrt{3}},\;T_{122}=\frac{\Delta^0}{\sqrt{3}},\;T_{222}=\Delta^-,\;
T_{113}=\frac{\Sigma^{*+}}{\sqrt{3}},\non
&&T_{123}=\frac{\Sigma^{*0}}{\sqrt{6}},\;T_{223}=\frac{\Sigma^{*-}}{\sqrt{3}},\;T_{133}=\frac{\Xi^{*0}}{\sqrt{3}},\;
T_{233}=\frac{\Xi^{*-}}{\sqrt{3}},\;T_{333}=\Omega^{-}.
\label{decuplet_field}
\ee
The phase convention that we follow implies the phases for the isospin states, $|\pi^+\rangle=-|1,1\rangle$, $|K^-\rangle=-|1/2,-1/2\rangle$, $|\Sigma^+\rangle=-|1,1\rangle$.

In Ref. \cite{Sarkar:2004jh} the expansion of the Lagrangian is done up to two mesons of incoming (outgoing) momentum $k(k\pr)$ which leads to an  interaction kernel of the form
\be
V_{ij}=-\frac{1}{4f^2}C_{ij}(k^0+k^{\pr 0})
\label{poten}
\ee 
for the $s$-wave transition amplitudes, as in Ref. \cite{kaon}.
For the quantum numbers strangeness $S=-1$ and isospin $I=0$ the relevant channels are
$\pi\Sgs$ and $K \Xi^*$ with the corresponding coefficients $C_{ij}$ given in Sec. \ref{sec:dwave}.

The matrix $V$ is then used as the kernel of the Bethe-Salpeter equation to
obtain the unitary transition matrix~\cite{kaon}. This results in the matrix equation
\be
T=(1-VG)^{-1}V
\label{BS}
\ee
where $G$ is a diagonal matrix representing the meson-baryon loop function
given in Ref. \cite{Sarkar:2005ap}. The loop function contains an
undetermined subtraction constant, which accounts for terms from higher order 
chiral Lagrangians that make it finite. In Ref. \cite{Sarkar:2005ap} the value of this constant has been fixed 
to $a_i=-2$ for a renormalization scale of $\mu=700$ MeV. However, once the $d$-wave channels are introduced in the 
coupled channel formalism, this constant will be allowed for fine tuning within close limits.

\subsection{Introduction of $d$-wave channels}
\label{sec:dwave}
As mentioned in the Introduction, a realistic coupled channel model for 
the $\Lambda^*(1520)$ should include also meson-baryon channels (MB) of the octet of $\pi$ with the octet of $p$ as the branching ratios
into $\overline{K}N$ and $\pi\Sigma$ are large. These latter states are then automatically in a $d$-wave state.
For the present study we include the $d$-wave channels following 
Ref. \cite{new_Oset_Sarkar_Roca}.
In a previous work \cite{Sarkar:2005ap} the $\Lambda^*(1520)$ resonance was
studied within a coupled channel formalism including the
$\pi\Sigma^*$, $K\Xi^*$ in $s$-wave and the $\bar K N$  and
$\pi\Sigma$ in $d$-waves leading to a good reproduction of the
pole position of the $\Lambda^*(1520)$ of the scattering
amplitudes. However, the use of the pole position to get the
properties of the resonance is far from being accurate as soon
as a threshold is opened close to the pole position on the real
axis, which is the present case with the $\pi\Sigma^*$ channel.

Apart from that, in the approach of Ref.~\cite{Sarkar:2005ap}
some matrix elements in the kernel of the Bethe-Salpeter
equation were not considered. 
Therefore, a subsequent work \cite{new_Oset_Sarkar_Roca} aimed at a
more  precise description of the physical processes involving
the $\Lambda^*(1520)$ resonance. Hence,  other possible
tree level transition potentials in $d$-wave are introduced here
following Ref. \cite{new_Oset_Sarkar_Roca}:
$\bar{K}N\to\bar{K}N$, $\bar{K}N\to\pi\Sigma$ and
$\pi\Sigma\to\pi\Sigma$. 
For these vertices, effective
transition potentials are used which are 
proportional to the incoming and outgoing momentum squared 
in order to account for the $d$-wave character of the
channels which will be formalized in the following. 
 
\begin{figure}[ht]
\centerline{
\includegraphics[width=0.4\textwidth]{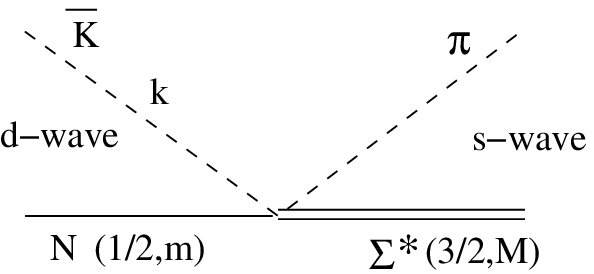}
}
\caption{The $\bk N\rw \pi\Sgs$ vertex}
\label{point}
\end{figure}

Consider the transition $\bk N$ ($d$-wave) to  $\pi\Sgs$ ($s$-wave) as shown in
Fig.~\ref{point}. 
We start with an amplitude of the form
\be
-it_{\bk N\rw\pi\Sgs}=-i\beta_{\bk N}\ |{\bf  k}|^2 \left[T^{(2)\dagger}
\otimes Y_{2}(\hat{{\bf k}})
\right]_{0\,0}
\ee
where $ T^{(2)\, \dagger}$ is a (rank 2) spin transition operator 
defined by
\[\langle 3/2\ M|\ T^{(2)\dagger}_\mu\ |1/2\ m\rangle
={\cal C}(1/2\ 2\
3/2;m \ \mu \ M)\ \langle 3/2||\ T^{(2) \dagger}\ ||1/2\rangle~,\]
$Y_2(\hat{{\bf  k}})$ is the spherical harmonic coupled to $ T^{(2) \dagger}$ to
produce a scalar, and
${\bf  k}$ is the momentum of the $\bk$. The third component of spin 
of the initial nucleon and the final $\Sgs$ are denoted by $m$
and $M$ respectively as indicated in the Clebsch-Gordan coefficients.
The coupling strength $\beta$ is not determined from theory and has to be fixed 
from experiment as has been done in Ref. \cite{new_Oset_Sarkar_Roca} 
with the results outlined below.
Choosing appropriately the reduced matrix element
we obtain 
\be
-it_{\bk N\rw\pi\Sgs}=-i\beta_{\bk N}\ |{\bf  k}|^2\ {\cal C}(1/2\ 2\
3/2;m,M-m)Y_{2,m-M}(\hat{{\bf  k}})(-1)^{M-m}\sqrt{4\pi}.
\ee
In the same way the amplitude for $\pi\Sigma$ ($d$-wave) to  $\pi\Sgs$ ($s$-wave) is written as 
\be
-it_{\pi\Sigma\rw\pi\Sgs}=-i\beta_{\pi\Sigma}\ |{\bf  k}|^2\ {\cal C}(1/2\ 2\
3/2;m,M-m)Y_{2,m-M}(\hat{{\bf  k}})(-1)^{M-m}\sqrt{4\pi}
\ee
and similarly for the rest of the transitions mentioned above.
The angular dependence disappears in the
loop integrations \cite{Sarkar:2005ap}.
The loop function of the meson-baryon system in $d$-wave is strongly 
divergent, but an on-shell factorization can be achieved \cite{Sarkar:2005ap} using arguments from the $N/D$ method 
from Ref. \cite{nsd} as explained in the former subsection.
The on-shell factorization ensures at the same time the unitarity of the amplitude
after solving the Bethe-Salpeter equation (\ref{BS}).

Denoting  the $\pi\Sigma^*$, $K\Xi^*$,  $\bar K N$,  and
$\pi\Sigma$ channels by $1$, $2$, $3$ and $4$, respectively, the
kernel $V$ of the Bethe-Salpeter equation (\ref{BS}) is written as:
\renewcommand{\arraystretch}{1.25}
\be
V=\left(
\begin{array}{cccc}
C_{11}(k_1^0+k_1^0)\ & C_{12}(k_1^0+k_2^0) & \gamma_{13}\,q_3^2 &\gamma_{14}\,q^2_4 \\
C_{21}(k_2^0+k_1^0)\ & C_{22}(k_2^0+k_2^0) & 0 & 0 \\
\gamma_{13}\,q_3^2 & 0 & \gamma_{33}\, q^4_3 & \gamma_{34} \,q_3^2 \,q^2_4\\
\gamma_{14}\,q^2_4  & 0 & \gamma_{34} \,q_3^2 \,q_4^2 &  \gamma_{44}\, q^4_4
\end{array}
\right)~,
\label{vij}
\ee
\noindent
with the on-shell CM momenta $q_i=\frac{1}{2\sqrt{s}}\sqrt{[s-(M_i+m_i)^2][s-(M_i-m_i)^2]}$, meson energy
$k_i^0=\frac{s-M_i^2+m_i^2}{2\sqrt{s}}$,
and baryon(meson) masses $M_i(m_i)$. The elements $V_{11}$, $V_{12}$, $V_{21}$, $V_{22}$ come from the
lowest order chiral Lagrangian involving the decuplet of baryons and
the octet of pseudoscalar mesons as discussed in Sec. \ref{sec:form}; see also
 Ref.~\cite{Sarkar:2004jh,lutz}.
The coefficients $C_{ij}$ obtained from Eq. (\ref{lag1}) are $C_{11}=\frac{-1}{f^2}$,
 $C_{21}=C_{12}=\frac{\sqrt{6}}{4f^2}$ and $C_{22}=\frac{-3}{4f^2}$,
where $f$ is $1.15f_\pi$, with $f_\pi$ ($=93\mev$) the pion decay constant,
which is an average between $f_\pi$ and
$f_K$ as was used in Ref.~\cite{kaon} in
the related problem of the dynamical 
generation of the $\Lambda(1405)$.

In the kernel $V$ we neglect the elements $V_{23}$ and $V_{24}$ which involve the
tree level interaction of the $K\Xi^*$ channel with the $d$-wave channels
because the $K\Xi^*$ threshold is far from the $\Ls$ mass and its role
in the resonance structure is far smaller than that of the
$\pi\Sigma^*$. This is also the reason why the $K\Xi$ channel in $d$-wave is completely ignored.

Summarizing, the parameters of the model are five $d$-wave coupling strengths $\gamma_{ij}$. 
Additionally, the subtraction constants can be fine-tuned around their natural values of $-2$ and $-8$ for $s$-wave loops and $d$-wave loops, respectively. 
The fit to $\overline{K}N\to\overline{K}N$ and $\overline{K}N\to\pi\Sigma$ data has been performed in Ref. \cite{new_Oset_Sarkar_Roca} and the results for the parameter values can be found there.

In the study of the radiative decay of the $\Lambda^*(1520)$ we will need only the coupling strengths of the resonance to its coupled channels at the resonance position \cite{new_Oset_Sarkar_Roca}. The effective $s$-wave ($d$-wave) couplings $g_{\Lambda^*MB^*}$ ($g_{\Lambda^*MB}$) are obtained by expanding the amplitude around the pole in a Laurent series. The residue is then identified with the coupling strength as described in Sec. \ref{sec:numres} and we display the result for the $g$'s in the isospin $I=0$ channel
from Ref. \cite{new_Oset_Sarkar_Roca} in Tab. \ref{tab:couplings}. 

\begin{table}[ht]
\caption{Coupling strength of the dynamically generated $\Lambda^*(1520)$ to ($MB^*$) in $s$-wave and ($MB$) in $d$-wave \cite{new_Oset_Sarkar_Roca}.}
\begin{center}
\begin{tabular*}{0.4\textwidth}{@{\extracolsep{\fill}}ll|ll}
\hline\hline $g_{\Lambda^*\pi\Sigma^*}$&$g_{\Lambda^*K\Xi^*}$&$g_{\Lambda^*\overline{K}N}$&$g_{\Lambda^*\pi\Sigma}$
\\
\rule[-2mm]{0mm}{8mm}$0.91$&$-0.29$&$-0.54$&$-0.45$
\\
\hline\hline
\end{tabular*}
\end{center}
\label{tab:couplings}
\end{table}

\section{Radiative decay}
\label{sec:raddecay}
\begin{figure}
\centerline{\includegraphics[width=0.3\textwidth]{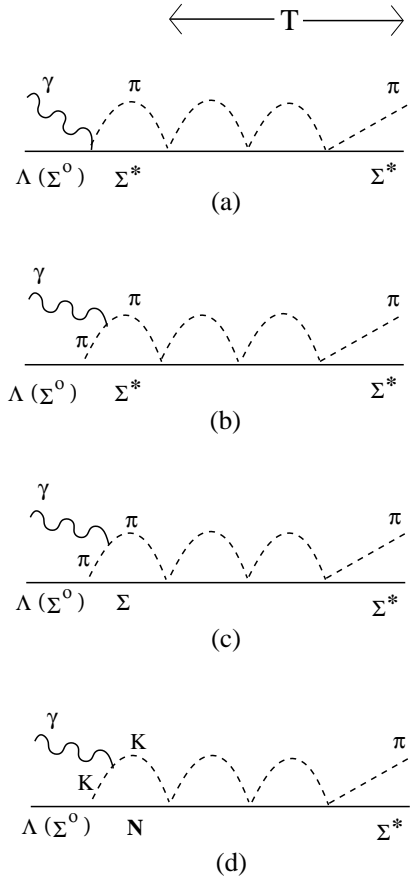}}
\caption{Coupling of the photon to the $\Lambda^*(1520)$. Diagrams (a) and (b) show the coupling to a $\pi\Sigma^*$ loop, which enters together with the corresponding diagrams in the $K\Xi^*$ channel. The rescattering series that generates the pole of the $\Lambda^*(1520)$ in the complex scattering plane is symbolized by $T$. Diagrams (c) and (d) show the $\gamma$ coupling to the $d$-waves of the resonance.}
\label{fig:mech1}
\end{figure}
For the radiative decay of the $\Lambda^*(1520)$ we study the reactions shown in Fig. \ref{fig:mech1} corresponding to $\gamma Y\to \pi\Sigma^*$. 
We consider in the loops all the meson-baryon states of the coupled channels and couple the photon to the first loop as shown in Fig. \ref{fig:mech1}.
In the loop attached to the photon we can have either $\pi\Sigma^*$ or $K\Xi^*$ that couples to the $\Lambda^*(1520)$ in $s$-wave or $\overline{K}N, \pi\Sigma$ which couple in $d$-wave. We show in the figure with the symbol $T$ the diagrams which are accounted by the $T(i\to\pi\Sigma^*)$ amplitude with $i$ any of the four channels $\pi\Sigma^*$, $K\Xi^*$, $\overline{K}N$, $\pi\Sigma$.
For the photon coupling we restrict ourselves to the Kroll-Ruderman (KR) and meson-pole (MP) coupling as shown in the figure. 
Formally, the photon should be also coupled to the meson and baryon components of the iteration of intermediate loops forming the $\Lambda^*(1520)$ but then the first loop vanishes for parity reasons ($p$-wave and $s$ or $d$-wave in the first loop). For the same reason the coupling of the photon to the $\Lambda(\Sigma^0)$ initial baryon would vanish. The coupling of the $\gamma$ to the baryon in the first loop vanishes in the heavy baryon limit and is very small otherwise. A general discussion of issues of gauge invariance, chiral invariance, etc., within the context of unitarized chiral theories can be found in \cite{Borasoy:2005zg,Doring:2005bx}. In Ref. \cite{Borasoy:2005zg} one proved that gauge invariance is preserved when the photon is coupled to internal as well as external lines and vertices. An extra discussion on this issue is given in \cite{Doring:2005bx}. According to these findings our present approach fulfills gauge invariance with errors of the order of 2\% from the approximations done.

For the diagrams from Fig. \ref{fig:mech1}, the $MBB^*$ vertices and the Kroll-Ruderman coupling $\gamma MBB^*$ are needed, for which we use the Lagrangian from Ref. \cite{Butler:1992pn}, with the part relevant for the present reaction given by
\be
{\cal L}={\cal C}\left(\overline{T}_\mu A^{\mu}B+{\overline B}A_\mu T^\mu\right)=
{\cal C}\left(\sum_{a,b,c,d,e}^{1,\cdots,3}\epsilon_{abc}\;\overline{T}^{ade} \;\overline{u}_\mu\;A_{d}^{b,\mu} \;B^c_e
+\sum_{a,b,c,d,e}^{1,\cdots,3}\epsilon^{abc}\;{\overline B}^e_c \;A^{d}_{b,\mu} \;T_{ade}\;u^\mu\right)
\label{precise_L}
\ee
with the same phase conventions as in Eq. (\ref{lag1}) and the spin and flavor structure as given in Ref. \cite{Sarkar:2004jh} and Eq.  (\ref{decuplet_field}).
In Eq. (\ref{precise_L}), the axial current is expanded up to one meson field, 
\be
A^\mu=\frac{i}{2}\left(\xi\partial^\mu\xi^\dagger-\xi^\dagger\partial^\mu\xi\right)\stackrel{{\rm one}\;\Phi}{\longrightarrow} 
\frac{\partial^\mu\Phi}{\sqrt{2}f_\pi}, \quad
\xi=\exp\left(\frac{i\Phi}{\sqrt{2}f_\pi}\right),
\ee
$\Phi, B,\overline{B}$ are the standard meson and baryon $SU(3)$ fields \cite{Gasser:1984gg}, and $f_\pi=93$ MeV. For the Kroll-Ruderman vertex $\gamma MBB^*$, we couple the photon by minimal substitution to Eq. (\ref{precise_L}).
The coupling strength ${\cal C}$ is determined from the $\Delta(1232)$ decay, 
\be
\frac{{\cal C}}{\sqrt{2}f_\pi}=\frac{f^*_{\Delta\pi N}}{m_\pi}
\ee
with $f^*_{\Delta\pi N}=2.13$. The $SU(3)$ breaking in the decuplet beyond that from the different masses is of the order of 30\% as a fit of Eq. (\ref{precise_L}) to the partial decay widths of $\Delta(1232)$, $\Sigma^*$, and $\Xi^*$ shows \cite{Butler:1992pn,angels_phi}. In the present study, we do not take this breaking into account in order to be consistent with the model for the dynamical generation of the $\Lambda^*(1520)$ where the $SU(3)$ breaking from other sources than mass differences is also neglected. 

From Eq. (\ref{precise_L}) and from the minimal coupling with the photon, Feynman rules for
$(\Lambda, \Sigma^0)$ $\to MB^*$,  $\gamma(\Lambda, \Sigma^0)$ $\to MB^*$, and the ordinary $\gamma MM$ vertices are obtained where the meson momentum ${\bf q}$ is defined as outgoing and the photon momentum ${\bf k}$ as incoming, 
\be
&&(-it)_{B\to M({\bf q})B^*}=\frac{d\;f^*_{\Delta\pi N}}{m_\pi}\; {\bf S}^\dagger\cdot {\bf q},
\quad 
\left(-i {\bf t}\cdot\boldsymbol{\epsilon}\right)_{KR}=-\;\frac{e\;c\;d\;f^*_{\Delta\pi N}}{m_\pi}\;{\bf S}^\dagger\cdot\boldsymbol{\epsilon},
\non 
&&\left(-i {\bf t}\cdot\boldsymbol{\epsilon}\right)_{\gamma({\bf k})M({\bf q}-{\bf k})\to M({\bf q})}=iec(2{\bf q}-{\bf k})\cdot\boldsymbol{\epsilon},
\non
\label{rule_decuplet}
\ee
with the coefficients $d$ given in Tab. \ref{tab:decuplet1}. 
\begin{table}
\caption{Coefficients $d$ for the Feynman rule Eq. (\ref{rule_decuplet}) with $\Lambda$ or $\Sigma^0$ in initial state.}
\begin{center}
\begin{tabular*}{0.7\textwidth}{@{\extracolsep{\fill}}llll}
\hline\hline &$\pi^-\Sigma^{*+}$&$\pi^+\Sigma^{*-}$&$K^+\Xi^{*-}$
\\
\hline
\rule[-4mm]{0mm}{10mm}$d,\;\Lambda\to MB^*$&$-\frac{1}{\sqrt{2}}$&
$\frac{1}{\sqrt{2}}$&$\frac{1}{\sqrt{2}}$
\\
\rule[-4mm]{0mm}{9mm}$d,\;\Sigma^0\to MB^*$&$-\frac{1}{\sqrt{6}}$&$-\frac{1}{\sqrt{6}}$
&$-\frac{1}{\sqrt{6}}$
\\
\hline\hline
\end{tabular*}
\end{center}
\label{tab:decuplet1}
\end{table}
In Eq. (\ref{rule_decuplet}) $e>0$ is the electron charge and $c=+1$ ($c=-1$) for $\pi^+$, $K^+$ ($\pi^-$, $K^-$) and $c=0$ for processes with neutral mesons. The photon with the polarization $\epsilon^\mu$ is real and we use the Coulomb gauge $\epsilon^0=0$, $\boldsymbol{ \epsilon}\cdot {\bf k}=0$.

For the first diagram in Fig. \ref{fig:mech1} in which $\pi^-\Sigma^{*+}$, $\pi^+\Sigma^{*-}$, $K^+\Xi^{*-}$ couple in $s$-wave to $T$, we construct the amplitude for the reactions $\gamma\Lambda\to\pi\Sigma^*$ and $\gamma\Sigma\to\pi\Sigma^*$ with isospin $I=0$. For this purpose, an isospin combination for the first loop is constructed
according to
\be
|\pi\Sigma^*, I=0\rangle &=&-\frac{1}{\sqrt{3}}|\pi^+\Sigma^{*-}\rangle-\frac{1}{\sqrt{3}}|\pi^0\Sigma^{*0}\rangle+\frac{1}{\sqrt{3}}|\pi^-\Sigma^{*+}\rangle,\non
|K\Xi^*, I=0\rangle &=& \frac{1}{\sqrt{2}}|K^+\Xi^{*-}\rangle-\frac{1}{\sqrt{2}}|K^0\Xi^{*0}\rangle
\label{isocombbstar}
\ee
with the phase conventions from above. Note that states with neutral mesons do not contribute to the loops. Using the Feynman rules from Eq. (\ref{rule_decuplet}), the results are [indicating, e.g., $\pi\Sigma^*$ in the first loop by $(\pi\Sigma^*)$]
\be
\left(-i {\bf t}\cdot\boldsymbol{\epsilon}\right)_{\rm KR}^{(I=0)}[\gamma\Lambda\to(\pi\Sigma^*)
\stackrel{{\Lambda^*}}{\to}\pi\Sigma^*]&=&0,\non
\left(-i {\bf t}\cdot\boldsymbol{\epsilon}\right)_{\rm KR}^{(I=0)}[\gamma\Lambda\to(K\Xi^*)\stackrel{{\Lambda^*}}{\to}\pi\Sigma^*]&=&
-\frac{e}{2}\frac{f^*_{\Delta\pi N}}{m_\pi}\;G_{2}\;T^{(21)}\;{\bf S}^\dagger\cdot\boldsymbol{\epsilon},\non
\left(-i {\bf t}\cdot\boldsymbol{\epsilon}\right)_{\rm KR}^{(I=0)}[\gamma\Sigma^0\to(\pi\Sigma^*)\stackrel{{\Lambda^*}}{\to}\pi\Sigma^*]&=&
-\frac{\sqrt{2}e}{3}\frac{f^*_{\Delta\pi N}}{m_\pi}\;G_{1}\;T^{(11)}\;{\bf S}^\dagger\cdot\boldsymbol{\epsilon},\non
\left(-i {\bf t}\cdot\boldsymbol{\epsilon}\right)_{\rm KR}^{(I=0)}[\gamma\Sigma^0\to(K\Xi^*)\stackrel{{\Lambda^*}}{\to}\pi\Sigma^*]&=&
\frac{e}{2\sqrt{3}}\frac{f^*_{\Delta\pi N}}{m_\pi}\;G_{2}\;T^{(21)}\;{\bf S}^\dagger\cdot\boldsymbol{\epsilon}
\label{amplbstar}
\ee
with $T^{(ij)}$ being the matrix element obtained from the Bethe-Salpeter equation (\ref{BS}) with the channel ordering $(ij)$ as in Eq. (\ref{vij}).
In Eq. (\ref{amplbstar}), $G_1$ and $G_2$ are the ordinary loop functions for $\pi\Sigma^*$ and $K\Xi^*$ given by
\be
G_i=\int\frac{d^3{\bf q}}{\left(2\pi\right)^3}\;\frac{1}{2\omega}\;\frac{1}{\sqrt{s}-\omega({\bf q})-E({\bf q})+i\epsilon}
\label{krphoton}
\ee
with the total CM energy $\sqrt{s}$, meson and baryon energy $\omega$ and $E$.  For the regularization a cut-off $\Lambda$ is used.
This cut off is determined such that the $G_i$ functions of Eq. (\ref{krphoton}) have the same value as obtained in
\cite{new_Oset_Sarkar_Roca} using dimensional regularization. For this purpose we match the $MB^*$ loop function in both regularization schemes (dimensional and cut-off) at $s^{1/2}=1520$ MeV which results in  $\Lambda_{\pi\Sigma^*}=418$ MeV for the $\pi\Sigma^*$ channel. This value is then used as the cut-off for Eq. (\ref{krphoton}). For the $K\Xi^*$ channel such a matching is not possible at energies so far below the $K\Xi^*$ threshold, and we set $\Lambda_{K\Xi^*}=500$ MeV. In any case, the final numbers are almost independent of the value of $\Lambda_{K\Xi^*}$, first, because the contribution is tiny and, second, because the cut-off dependence of the $s$-wave loops is moderate. 

In order to evaluate the contribution of the meson-pole term in the second diagram of Fig. \ref{fig:mech1}, we must project the operator $\boldsymbol{\epsilon}\cdot(2{\bf q}-{\bf k})\;{\bf S}^\dagger\cdot ({\bf q-k})$ onto $s$-wave; for this we neglect ${\bf k}$ which is relatively small in the radiative decay (the numerical test keeping the ${\bf k}$ terms proves this to be a very good approximation). Then, we get as a projection ${\bf S}^\dagger\cdot \boldsymbol{\epsilon}\;\frac{2}{3}\;{\bf q}^2$ and we have a new loop function
\be
\tilde{G}_i&=&i\;\int\frac{d^4 q}{\left(2\pi\right)^4}\;\frac{{\bf q}^2}{(q-k)^2-m_i^2+i\epsilon}\;\frac{1}{q^2-m_i^2+i\epsilon}\;\frac{1}{P^0-q^0-E_i({\bf q})+i\epsilon},\non
&=&-\int\frac{d^3{\bf q}}{\left(2\pi\right)^3}\;\frac{{\bf q}^2}{2\omega_i\omega_i'}\;\frac{1}{k+\omega_i+\omega_i'}\;\frac{1}{k-\omega_i-\omega_i'+i\epsilon}\;
\frac{1}{\sqrt{s}-\omega_i-E_i({\bf q})+i\epsilon}\;\frac{1}{\sqrt{s}-k-\omega_i'-E_i({\bf q})+i\epsilon},\non
&&\left[\left(\omega_i+\omega_i'\right)^2+\left(\omega_i+\omega_i'\right)\left(E_i({\bf q})-\sqrt{s}\right)+k\omega_i'\right]
\label{tildeg}
\ee
where $\omega_i$ and $\omega_i'$ are the energies of the mesons of mass $m_i$ at momentum ${\bf q}$ and ${\bf q-k}$, respectively, $k$ is the energy of the on-shell photon and $E_i$ the energy of the decuplet baryon. 
For the regularization of the loop we use the same cut-offs as for Eq. (\ref{krphoton}) from above.
The diagrams with meson-pole terms can be easily incorporated by changing
$G_i\to G_i+\frac{2}{3}\;{\tilde G}_i$ in Eq. (\ref{amplbstar}), resulting in 
\be
\left(-i {\bf t}\cdot\boldsymbol{\epsilon}\right)_{\rm KR+MP}^{(I=0)}[\gamma\Lambda\to(\pi\Sigma^*)\stackrel{{\Lambda^*}}{\to}\pi\Sigma^*]&=&0,\non
\left(-i {\bf t}\cdot\boldsymbol{\epsilon}\right)_{\rm KR+MP}^{I=0}[\gamma\Lambda\to(K\Xi^*)\stackrel{{\Lambda^*}}{\to}\pi\Sigma^*]&=&
-\frac{e}{2}\frac{f^*_{\Delta\pi N}}{m_\pi}\left(G_{2}+\frac{2}{3}\;\tilde{G}_2\right)\;T^{(21)}\;{\bf S}^\dagger\cdot\boldsymbol{\epsilon},\non
\left(-i {\bf t}\cdot\boldsymbol{\epsilon}\right)_{\rm KR+MP}^{(I=0)}[\gamma\Sigma^0\to(\pi\Sigma^*)\stackrel{{\Lambda^*}}{\to}\pi\Sigma^*]&=&
-\frac{\sqrt{2}e}{3}\frac{f^*_{\Delta\pi N}}{m_\pi}\;\left(G_{1}+\frac{2}{3}\;\tilde{G}_1\right)\;T^{(11)}\;{\bf S}^\dagger\cdot\boldsymbol{\epsilon},\non
\left(-i {\bf t}\cdot\boldsymbol{\epsilon}\right)_{\rm KR+MP}^{(I=0)}[\gamma\Sigma^0\to(K\Xi^*)\stackrel{{\Lambda^*}}{\to}\pi\Sigma^*]&=&
\frac{e}{2\sqrt{3}}\frac{f^*_{\Delta\pi N}}{m_\pi}\;\left(G_{2}+\frac{2}{3}\;\tilde{G}_2\right)\;T^{(21)}\;{\bf S}^\dagger\cdot\boldsymbol{\epsilon}.
\label{amplswave}
\ee

\subsection{Radiative decay from $d$-wave loops}
\label{sec:raddwave2}
The third and fourth diagram in Fig. \ref{fig:mech1} show the photon coupling to the $d$-wave components of the $\Lambda^*(1520)$. The first loop implies two $p$-wave and one $d$-wave couplings which lead to a non-trivial angular momentum structure. Note that there is no coupling of the Kroll-Ruderman type because the combination of $s$ and $d$-wave couplings vanishes by parity in the loop integration. 

The $MBB$ $p$-wave coupling is obtained from the lowest order chiral meson-baryon Lagrangian \cite{Gasser:1984gg} which leads to the Feynman rule (meson momentum ${\bf p}$ outgoing)
\be
(-it)=i{\cal L}=-\frac{\sqrt{2}}{f_\pi}\;\boldsymbol{\sigma}\cdot {\bf p}\left(a\;\frac{D+F}{2}+b\;\frac{D-F}{2}\right)
\label{feynman_BMM}
\ee
with $a$ and $b$ given in Tab. \ref{tab:feynman_1} where only the channels including charged mesons are denoted.
\begin{table}
\caption{Coefficients $a$ and $b$ for the Feynman rule Eq. (\ref{feynman_BMM}), meson momentum {\it outgoing}.}
\begin{center}
\begin{tabular*}{0.5\textwidth}{@{\extracolsep{\fill}}llllll}
\hline\hline &$\pi^-\Sigma^+$&$\pi^+\Sigma^-$&$K^-p$
\\
\hline
\rule[-4mm]{0mm}{10mm}$a,\;\Lambda\to MB$&$\frac{1}{\sqrt{6}}$&
$\frac{1}{\sqrt{6}}$&$-\sqrt{\frac{2}{3}}$
\\
\rule[-4mm]{0mm}{9mm}$b,\;\Lambda\to MB$&$\frac{1}{\sqrt{6}}$&$\frac{1}{\sqrt{6}}$
&$\frac{1}{\sqrt{6}}$
\\
\rule[-4mm]{0mm}{9mm}$a,\;\Sigma^0\to MB$&$-\frac{1}{\sqrt{2}}$&$\frac{1}{\sqrt{2}}$&$0$
\\
\rule[-4mm]{0mm}{9mm}$b,\;\Sigma^0\to MB$&$\frac{1}{\sqrt{2}}$&$-\frac{1}{\sqrt{2}}$&$\frac{1}{\sqrt{2}}$
\\
\hline\hline
\end{tabular*}
\end{center}
\label{tab:feynman_1}
\end{table}
As in the last section, the isospin zero channel is constructed from the particle channels according to 
\be
|\pi\Sigma, I=0\rangle&=&-\frac{1}{\sqrt{3}}\;|\pi^+\Sigma^-\rangle -\frac{1}{\sqrt{3}}\;|\pi^0\Sigma^0\rangle-\frac{1}{\sqrt{3}}\;|\pi^-\Sigma^+\rangle,\non
|\overline{K} N, I=0\rangle&=&\frac{1}{\sqrt{2}}\;|\overline{K}^0 n\rangle+\frac{1}{\sqrt{2}}\;|K^- p\rangle .
\label{isocombb}
\ee
Using the Feynman rules from Eq. (\ref{feynman_BMM}) and from Eq. (\ref{rule_decuplet}) for the $\gamma MM$ vertex, the amplitudes read 
\be
\left(-i {\bf t}\cdot\boldsymbol{\epsilon}\right)^{(I=0)}[\gamma\Lambda\to(\pi\Sigma)\stackrel{{\Lambda^*}}{\to}\pi\Sigma^*]&=&0,\non
\left(-i {\bf t}\cdot\boldsymbol{\epsilon}\right)^{(I=0)}[\gamma\Lambda\to({\overline K}N)\stackrel{{\Lambda^*}}{\to}\pi\Sigma^*]&=&
\frac{e}{\sqrt{2}f_\pi}\left(\frac{D}{3}\;+F\right)\;{\tilde G}'_3 T^{(31)}\;{\bf S}^\dagger\cdot\boldsymbol{\epsilon},\non
\left(-i {\bf t}\cdot\boldsymbol{\epsilon}\right)^{(I=0)}[\gamma\Sigma^0\to(\pi\Sigma)\stackrel{{\Lambda^*}}{\to}\pi\Sigma^*]&=&
-\frac{4eF}{3f_\pi}\;{\tilde G}'_4 T^{(41)}\;{\bf S}^\dagger\cdot\boldsymbol{\epsilon},\non
\left(-i {\bf t}\cdot\boldsymbol{\epsilon}\right)^{(I=0)}[\gamma\Sigma^0\to({\overline K}N)\stackrel{{\Lambda^*}}{\to}\pi\Sigma^*]&=&
\frac{e}{\sqrt{6}f_\pi}\;\left(F-D\right)\;{\tilde G}'_3 T^{(31)}\;{\bf S}^\dagger\cdot\boldsymbol{\epsilon}
\label{ampldwave}
\ee
with the channel ordering $i=1,\cdots, 4$ being $\pi\Sigma^*$, $K\Xi^*$, ${\overline K}N$, $\pi\Sigma$ as in the last sections.
As above, we have chosen $\pi\Sigma^*$ as the final state which will become clear in Sec. \ref{sec:numres} when the coupled channel scheme is matched with a formalism with explicit excitation of the resonance. 

The loop function ${\tilde G}'_i$ in Eq. (\ref{ampldwave}) for the first loop is given by
\be
{\tilde G}'_i&=&i\;\int\frac{d^4 q}{\left(2\pi\right)^4}\;\frac{{\bf q}^2}{(q-k)^2-m_i^2+i\epsilon}\;\frac{1}{q^2-m_i^2+i\epsilon}\;\frac{1}{P^0-q^0-E_i({\bf q})+i\epsilon}\;\frac{M}{E_i({\bf q})}\left(\frac{{\bf q}^2}{q_{\rm on}^2}\right)
\label{gtildep}
\ee
which is similar to ${\tilde G}$ from Eq. (\ref{tildeg}) up to a factor $M/E$ from the non-relativistic reduction of the baryon propagator and a factor ${\bf q}^2/q_{\rm on}^2$. 
As in the case of the $MB^*$ $s$-wave loops, the divergence in Eq. (\ref{gtildep}) is regularized by a cut-off whose value is obtained by matching dimensional regularization and cut-off scheme of the meson-baryon $d$-wave loop at $s^{1/2}=1520$ MeV as explained following Eq. (\ref{krphoton}). With the subtraction constant from Ref. \cite{Sarkar:2005ap,new_Oset_Sarkar_Roca}, values for the cut-off of $\Lambda_{\overline{K}N}=507$ MeV and $\Lambda_{\pi\Sigma}=558$ MeV follow. 
In the following subsection we present the technical details which have led to Eqs. (\ref{ampldwave}) and (\ref{gtildep}), projecting the meson-pole term over $d$-waves and performing the angular integrations.

\subsubsection{The spin-polarization structure of $d$-wave loops}
The structure of the two $p$-wave couplings of the first loop in the fourth diagram of Fig. \ref{fig:mech1} is given by
\be
\eps^\mu(2q-k)_\mu\ \boldsymbol{\sigma}\cdot({\bf  k}-\bf{ q})
\ee
where the meson momentum of the $MBB$ vertex is given by $q-k$ and the two mesons in the $\gamma MM$ vertex are at momentum $q-k$ and $q$.
As $\epsilon^0=0$ in Coulomb gauge, the spin structure takes the form $\boldsymbol\eps\cdot\boldsymbol q\ \boldsymbol\sigma\cdot \bf q$ (neglecting the photon momentum $\boldsymbol k$ which is small in the radiative decay). The $d$-wave structure obtained from $\sigma_iq_i\eps_jq_j\to\sigma_i\eps_j(q_iq_j-\tfrac{1}{3}{\bf q}^2\delta_{ij})$ will combine with the $d$-wave structure $Y_2(\hat{\bf q})$ coming from the $\bk N\to\pi\Sgs$ vertex to produce a scalar quantity after the loop integration is performed (for the second loop, we choose the $\pi\Sgs$ channel in the following, but the calculations hold for any of the four channels in the second loop).

We write
\be
\sigma_i\eps_j(q_iq_j-\tfrac{1}{3}{\bf q}^2\delta_{ij})
=A\left[[\sigma\otimes\eps]_\mu^2\ Y_2(\hat{\bf q})\right]_0^0
\label{eq:dwave}
\ee  
which indicates that the two vector operators $\vec{\sigma}$ and $\vec{\eps}$ couple to produce an operator of rank $2$ which couples to the spherical harmonic $Y_2(\hat{\bf q})$ to produce a scalar. The right-hand side can be written as
\be
&&A\sum_\mu(-1)^\mu [\sigma\otimes\eps]_\mu^2\ Y_{2,-\mu}(\hat{\bf q})=A\sum_{\mu,\alpha}(-1)^\mu Y_{2,-\mu}(\hat{\bf q})\ \C(1\ 1\ 2;\alpha,\mu-\alpha)\sigma_\alpha\eps_{\mu-\alpha}
\ee
where $\C$ denotes the Clebsch Gordan coefficient.
To find the value of $A$ we take the matrix element of both sides of Eq.~(\ref{eq:dwave}) between the states $m$ and $m'$ so that
\be
\langle m|\sigma_i\eps_j(q_iq_j-\tfrac{1}{3} {\bf q}^2\delta_{ij})|m'\rangle
&=&A\sum_\mu(-1)^\mu\ Y_{2,-\mu}(\hat{\bf q})\ \eps_{\mu-m+m'}\nonumber\\
&\times&\C(1\ 1\ 2;m-m',\mu-m+m')\ \C(\tfrac{1}{2}\ 1\ \tfrac{1}{2};m',m-m')
\label{eq:dwave1}
\ee
where we have used $\langle m|\sigma_\alpha|m'\rangle=\sqrt{3}\ \C(\tfrac{1}{2}\ 1\ \tfrac{1}{2};m',\alpha,m)$. Taking specific values of spin $1/2$ components, $m$ and $m'$, we obtain
\be
A=\sqrt{\frac{8\pi}{15}}\;{\bf q}^2~.
\label{eq:a}
\ee
Following Ref.~\cite{Sarkar:2005ap}, we now include the $\bk N\to\pi\Sgs$ vertex given by
\be
-it_{\bk N\to\pi\Sgs}=-i\beta_{\bk N}\ |{\bf q}|^2\ \C(\tfrac{1}{2}\ 2\
\tfrac{3}{2};m,M-m)Y_{2,m-M}(\hat{\bf q})(-1)^{M-m}\sqrt{4\pi}
\ee
so that the total spin structure of the $d$-wave loop in Fig.~\ref{fig:mech1} is essentially given by
\be
J=\sum_m\int\frac{d\Omega_q}{4\pi}\langle m|\sigma_i\eps_j(q_iq_j-\tfrac{1}{3}{\bf q}^2\delta_{ij})|m'\rangle\ \C(\tfrac{1}{2}\ 2\ \tfrac{3}{2};m,M-m)Y_{2,m-M}(\hat{\bf q})(-1)^{M-m}\sqrt{4\pi}
\ee
where we have performed an average over the angles in the integration over the loop momentum $q$.
Using Eqs.~(\ref{eq:dwave1}) and (\ref{eq:a}) this can be written as
\be
J&=&\sqrt{\frac{2}{3}}\ {\bf q}^2\ (-1)^{1-M+m'}\ \eps_{m'-M}\nonumber\\
&&\times\sum_m\
\C(\tfrac{1}{2}\ 1\ \tfrac{1}{2};m',m-m')\ \C(\tfrac{1}{2}\ 2\ \tfrac{3}{2};m,M-m)
\ \C(1\ 2\ 1;m-m',M-m)
\ee
where we have used the well-known relations
\[\int\ d\Omega_q\ Y_{2,-\mu}(\hat{\bf q})\ Y_{2,m-M}(\hat{\bf q})=(-1)^\mu\delta_{\mu,m-M}\]
and
\[\C(1\ 1\ 2;m-m',m'-M)=(-1)^{1-m+m'}\sqrt{\tfrac{5}{3}}\ \C(1\ 2\ 1;m-m',M-m)~.
\]
The product of three Clebsch-Gordan coefficients is then combined into a single one with a Racah coefficient, resulting in the identity
\be
&&\sum_m\
\C(\tfrac{1}{2}\ 1\ \tfrac{1}{2};m',m-m')\ \C(\tfrac{1}{2}\ 2\ \tfrac{3}{2};m,M-m)
\ \C(1\ 2\ 1;m-m',M-m)\nonumber\\
&&=-\sqrt{\tfrac{1}{2}}\ \C(\tfrac{1}{2}\ 1\ \tfrac{3}{2};m',M-m')
\ee
so that we finally have
\be
J=\frac{1}{\sqrt{3}}\ {\bf q}^2\ {\bf S}^\dagger\cdot\boldsymbol{\eps}~.
\label{eq:vert1}
\ee
The above relation implies that for practical purposes we can use for the $d$-wave projection of the two $p$-wave vertices the simple form $\tfrac{1}{\sqrt{3}}\ {\bf q}^2\ {\bf S}^\dagger\cdot\boldsymbol{\eps}$ and for the the $d$-wave vertex of the $MB\to MB^*$ amplitude the factor $\beta_{\bk N}{\bf q}^2$ and continue with the formalism exactly as in $s$-wave. 

In the on-shell reduction scheme for the $d$-wave transitions in the generation of the $\Lambda^*$, the factor $q_{\rm on}^2$ from the vertex is absorbed in the kernel $V$ as can be seen in Eq. (\ref{vij}). As we cannot perform this factorization for the first loop, we continue using the factor $\beta_{\bk N}{\bf q}^2$ for the $d$-wave vertex in this loop but then have to divide by $q_{\rm on}^2$ which will cancel the $q_{\rm on}^2$ in $V$ or the $T$ matrix.
All these factors considered, we obtain Eq. (\ref{ampldwave}) with ${\tilde G}'_i$ given in Eq. (\ref{gtildep}).

\section{Numerical results}
\label{sec:numres}
\begin{figure}
\centerline{\includegraphics[width=0.3\textwidth]{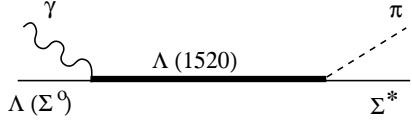}}
\caption{Effective resonance representation of the radiative decay.}
\label{fig:mech2}
\end{figure}
In the previous sections the amplitudes for the process $\gamma\Lambda\stackrel{{\Lambda^*}}{\to} \pi\Sigma^*$ and $\gamma\Sigma^0\stackrel{{\Lambda^*}}{\to} \pi\Sigma^*$ have been determined and are written in terms of the $T^{(i1)}$, the unitary solution of the Bethe-Salpeter equation (\ref{BS}) for meson-baryon scattering with the transitions from channel $i$ to the $\pi\Sigma^*$ final state. In order to determine the partial photon decay widths of the $\Lambda^*(1520)$, the $T^{(i1)}$ is expanded around the pole in the complex scattering plane and can be written as
\be
T^{(i1)}=\frac{g_i g_{\pi\Sigma^*}}{\sqrt{s}-M_{\Lambda^*(1520)}}.
\ee
The matrix elements from Eq. (\ref{amplswave}) and (\ref{ampldwave}) with this replacement for $T^{(i1)}$ is now identified with the resonant process in Fig. \ref{fig:mech2}, which is written as
\be
\left(-i {\bf t}\cdot\boldsymbol{\epsilon}\right)=\left(-ig_{\Lambda^*\pi\Sigma^*}\right)\;\frac{i}{\sqrt{s}-M_{\Lambda^*}}
\;g_{\Lambda^*\gamma\Lambda(\Sigma^0)}\;{\bf S}^\dagger\cdot\boldsymbol{\epsilon}.
\ee
This identification allows us to write the effective $\Lambda^*\gamma\Lambda$ and $\Lambda^*\gamma\Sigma^0$ couplings, $g_{\Lambda^*\gamma\Lambda}$ and $g_{\Lambda^*\gamma\Sigma^0}$, in terms of the couplings $g_{i1}$ of the $\Lambda^*(1520)$ in the transition of the channel $i\to\Lambda^*(1520)\to\pi\Sigma^*$ with its values given in Tab. \ref{tab:couplings}, resulting in
\be
g_{\Lambda^*\gamma\Lambda}^{(K\Xi^*)}&=&-\frac{e}{2}\frac{f_{\pi N\Delta}^*}{m_\pi}\;\left(G_2+\frac{2}{3}\;\tilde{G}_2\right)g_{\Lambda^* K\Xi^*},\non
g_{\Lambda^*\gamma\Sigma^0}^{(\pi\Sigma^*)}&=&-\frac{\sqrt{2}e}{3}\frac{f_{\pi N\Delta}^*}{m_\pi}\;\left(G_1+\frac{2}{3}\;\tilde{G}_1\right)g_{\Lambda^* \pi\Sigma^*},\non
g_{\Lambda^*\gamma\Sigma^0}^{(K\Xi^*)}&=&\frac{e}{2\sqrt{3}}\frac{f_{\pi N\Delta}^*}{m_\pi}\;\left(G_2+\frac{2}{3}\;\tilde{G}_2\right)g_{\Lambda^* K\Xi^*},\non
g_{\Lambda^*\gamma\Lambda}^{(\overline{K}N)}&=&\frac{e(D+3F)}{3\sqrt{2}f_\pi}\;\tilde{G}'_3\;g_{\Lambda^* \overline{K}N},\non
g_{\Lambda^*\gamma\Sigma^0}^{(\pi\Sigma)}&=&-\frac{4eF}{3f_\pi}\;\tilde{G}'_4\;g_{\Lambda^* \pi\Sigma},\non
g_{\Lambda^*\gamma\Sigma^0}^{(\overline{K}N)}&=&\frac{e(F-D)}{\sqrt{6}f_\pi}\;\tilde{G}'_3\;g_{\Lambda^* \overline{K}N}.
\ee
The upper index in brackets indicates which particles are present in the first loop. Adding all processes, we find using 
\be
g_{\Lambda^*\gamma\Lambda}&=&g_{\Lambda^*\gamma\Lambda}^{(K\Xi^*)}+g_{\Lambda^*\gamma\Lambda}^{(\overline{K}N)},\non
g_{\Lambda^*\gamma\Sigma^0}&=&g_{\Lambda^*\gamma\Sigma^0}^{(\pi\Sigma^*)}+ g_{\Lambda^*\gamma\Sigma^0}^{(K\Xi^*)}+ g_{\Lambda^*\gamma\Sigma^0}^{(\pi\Sigma)}+g_{\Lambda^*\gamma\Sigma^0}^{(\overline{K}N)},
\ee
the partial decay width for the processes $\Lambda^*(1520)\to\gamma\Lambda$ and $\Lambda^*(1520)\to\gamma\Sigma^0$ is given by
\be
\Gamma=\frac{k}{3\pi}\;\frac{M_Y}{M_{\Lambda^*}}\;|g_{\Lambda^*\gamma Y}|^2
\ee
where $Y=\Lambda,\;\Sigma^0$ is the final state hyperon and $k=\lambda^{1/2}(M_{\Lambda^*},0,M_Y^2)/(2M_{\Lambda^*})$ the CM momentum of the decay products.

In Tab. \ref{tab:numres} the numerical results from this study are compared with experimental data.
\begin{table}
\caption{Experimental data, quark model results from Ref. \cite{Kaxiras:1985zv}, and results from this study for the partial decay width of the $\Lambda^*(1520)$ into $\gamma\Lambda$ and $\gamma\Sigma^0$.}
\begin{center}
\begin{tabular*}{0.6\textwidth}{@{\extracolsep{\fill}}lll}
\hline\hline &$\Gamma\left(\Lambda^*(1520)\to\gamma\Lambda\right)$ [keV]&$\Gamma\left(\Lambda^*(1520)\to\gamma\Sigma^0\right)$ [keV]
\\
\hline
From Ref. \cite{Bertini:1987ye}&$33\pm 11$&$47\pm 17$
\\
From Ref. \cite{Mast:1969tx}&$134\pm 23$&
\\
From Ref. \cite{Antipov:2004qp}&$159\pm 33\pm 26$&
\\
From Ref. \cite{Taylor:2005zw}&$167\pm 43_{-12}^{+26}$&
\\
\hline
From Ref. \cite{Kaxiras:1985zv}&$46$&$17$
\\
\hline
This study&$3$&$71$
\\
\hline\hline
\end{tabular*}
\end{center}
\label{tab:numres}
\end{table}
For the $\gamma\Sigma^0$ final state, our result almost matches within errors the value given in Ref. \cite{Bertini:1987ye}, and certainly matches it considering the theoretical uncertainties that we will estimate below. 
The experimental value from Ref. \cite{Bertini:1987ye} is the only direct measurement of $\Gamma(\Lambda^*\to \gamma\Sigma^0)$. In the same experiment \cite{Bertini:1987ye}, the $\Gamma(\Lambda^*\to \gamma\Lambda)$ partial width has also been determined but lies far below more recent measurements, see Tab. \ref{tab:numres}. Note, that the value from Ref. \cite{Eidelman:2004wy} for $\Gamma(\Lambda^*\to \gamma\Sigma^0)$ is around six times larger than the value from Ref. \cite{Bertini:1987ye}. However, this large value is not a direct measurement (see Ref. \cite{Oh:2005ri}) but is extrapolated from $\Gamma(\Lambda^*\to \gamma\Lambda)$ by using $SU(3)$ arguments in Ref. \cite{Mast:1969tx}. Summarizing, the experimental situation is far from being clear. In the present study we compare to the direct measurement of $\Gamma(\Lambda^*\to \gamma\Sigma^0)=47\pm 17$ keV as a reference, but an independent experimental confirmation of this value would be desirable. Efforts in this direction have been announced \cite{Vavilov:2005sa}.

The theoretical value for the $\gamma\Lambda$ final state in Tab. \ref{tab:numres} is systematically below experiment although there are large discrepancies in the data. This suggests that the decay mechanisms could come from a different source than the coupled hadronic channels. 
The theoretical value is small because of large cancellations:  In the scheme of dynamical generation, the dominant building channel of the $\Lambda^*(1520)$ is given by $\pi\Sigma^*$  as can be seen in Tab. \ref{tab:couplings}. However, in the isospin combination from Eq. (\ref{isocombbstar}) which is needed in Eq. (\ref{amplswave}), this channel precisely vanishes because of the cancellation of the $\pi^+\Sigma^{*-}$ and $\pi^-\Sigma^{*+}$ contributions. The same holds for the $\pi\Sigma$ channel in $d$-wave with the cancellation in Eq. (\ref{ampldwave}) from the isospin combination in Eq. (\ref{isocombb}). This channel is important as the branching ratio into $\pi\Sigma$ is large. 
In contrast, the diagrams with $\pi^+\Sigma^{*-}$ and $\pi^-\Sigma^{*+}$ add in the $I=0$ combination with $\gamma\Sigma^0$ in the final state instead of $\gamma\Lambda$, as Eq. (\ref{amplswave}) shows, and the same is true for $\pi\Sigma$ in $d$-wave. As a result, a much larger partial decay width for the $\gamma\Sigma^0$ final state is obtained. 

\begin{figure}
\centerline{\includegraphics[width=0.6\textwidth]{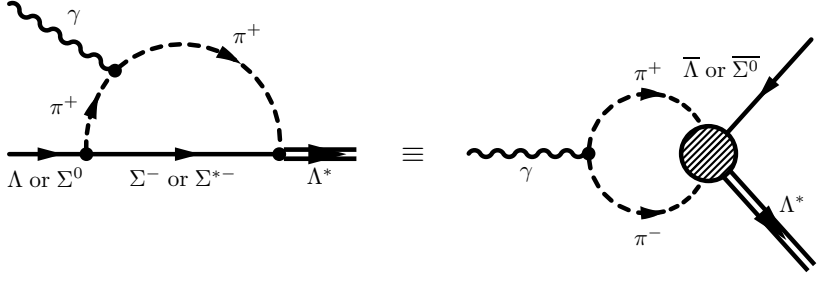}}
\caption{Alternative representation of the photonic loop with $\pi\Sigma$ and $\pi\Sigma^*$.}
\label{fig:turned}
\end{figure}

The cancellation of the $\pi\Sigma$ and $\pi\Sigma^*$ channels can be also understood when we turn the external baryon line around and redraw the decay process as shown in Fig. \ref{fig:turned}.
First, we consider the case with the $\Lambda$. The $\pi^+\pi^-$ system is necessarily in $J^P=1^-$ as these are the quantum numbers of the photon. As a consequence, the condition $L+S+I=$ {\it even} for the two-pion state where $L=J=1$ and $S=0$ can only be fulfilled if the two-pion state is in $I=1$; this is in contradiction to $I=0$ of the $\overline{\Lambda}\Lambda^*$ system. This is independent of the interaction denoted with the gray dashed circle in Fig. \ref{fig:turned}. In contrast, if the baryon on the right side is a $\overline{\Sigma^0}$, then the $\overline{\Sigma^0}\Lambda^*$ system is in an isospin one state, so that a finite contribution is expected. If the $\pi^+\pi^-$ system is replaced with $K^+K^-$, there is no restriction imposed by $L+S+I=$ {\it even}, so this process is possible for both $\Lambda$ or $\overline{\Sigma^0}$ on the right side.

\begin{figure}
\centerline{\includegraphics[width=0.4\textwidth]{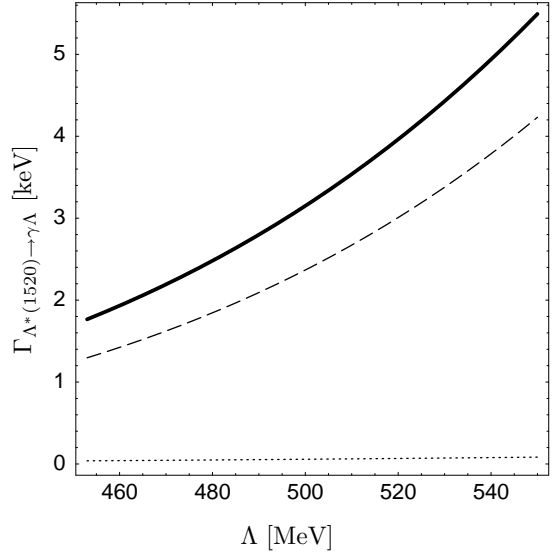}}
\caption{Cut-off dependency of $\Gamma\left(\Lambda^*(1520)\to\gamma\Lambda\right)$ [keV]. Contributions for different particles in the first loop and coherent sum. Dotted line: $K\Xi^*$ in $s$-wave. Dashed line: $\overline{K} N$ in $d$-wave. Thick solid line: Coherent sum.}
\label{fig:cutlambda}
\end{figure}

\begin{figure}
\centerline{\includegraphics[width=0.4\textwidth]{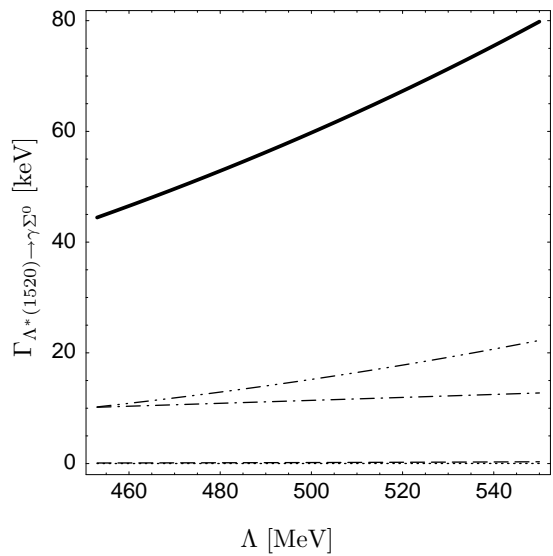}}
\caption{Cut-off dependency of $\Gamma\left(\Lambda^*(1520)\to\gamma\Sigma^0\right)$ [keV]. Contributions for different particles in the first loop and coherent sum. Dotted line: $K\Xi^*$ in $s$-wave. Dashed line: $\overline{K} N$ in $d$-wave. Dashed dotted line: $\pi\Sigma^*$ in $s$-wave. Double dashed dotted line: $\pi\Sigma$ in $d$-wave. Thick solid line: Coherent sum.}
\label{fig:cutsigma}
\end{figure}

The situation is illustrated in Fig. \ref{fig:cutlambda} and \ref{fig:cutsigma} where the partial decay widths are plotted as a function of the cut-off in the first loops.
Indeed, the large $\pi\Sigma$ and $\pi\Sigma^*$ channels that contribute in Fig. \ref{fig:cutsigma} are missing in Fig. \ref{fig:cutlambda} and render the width small. Note also that the $d$-wave loops introduce a relatively strong cut-off dependence. Our cut-offs from Secs. \ref{sec:raddecay} and \ref{sec:raddwave2} have been uniquely fixed by matching the cut-off scheme to the dimensional regularization scheme of the $MB^*$ and $MB$ loop functions that generate dynamically the $\Lambda^*(1520)$. The latter have values for the subtraction constants which lead to good data description in $\overline{K}N\to\overline{K}N$ and $\overline{K}N\to\pi\Sigma$ \cite{new_Oset_Sarkar_Roca}. Therefore, assuming that the strong interaction in these processes fixes the cut-offs, their values should be taken seriously and not changed for the first loop with the photon. On the other hand, the strong cut-off dependence is a large source of theoretical error in the model of the radiative decay such that uncertainties as big as 50 \% would not be exaggerated. With this uncertainty the $\Lambda^*(1520)\to \gamma\Sigma^0$ is clearly compatible with the only data available. But the $\Lambda^*(1520)\to \gamma\Lambda$ is certainly not. However, the fact that the only measurement for $\Lambda^*(1520)\to \gamma\Sigma^0$ is done in an experiment where the $\Lambda^*(1520)\to \gamma\Lambda$ disagrees so strongly with other measurements calls for caution and and further data on this decay rate is most needed.

On the other hand, even with large uncertainties our prediction for $\Lambda^*(1520)\to \gamma\Lambda$ is definitely small. Hence we have pinned down an observable which is extremely sensitive to extra components of the $\Lambda^*(1520)$ resonance beyond the meson-baryon ones provided by the chiral unitary approach. The sensitivity shows up because of the exact cancellation of the contribution from the most important components provided by the chiral unitary approach.

\section{Conclusions}
The chiral unitary model for the $\Lambda^*(1520)$ has been extended in order to describe the radiative decay of the $\Lambda^*$. 
The study of the two decay modes into $\gamma\Lambda$ and $\gamma\Sigma^0$ can help gain insight into the nature of the $\Lambda^*$, as to whether it is a genuine three quark state, a dynamically generated resonance, or a mixture of both. 

For the $\gamma\Sigma^0$ final state we have seen that the model of dynamical generation matches the empirical value, although there are certain theoretical uncertainties from the $d$-wave loops in the model. However, the good reproduction of the empirical value fits in the picture because the dominant channels of our coupled channel model add up for this decay, and in some quark models, the dominant three quark component for this decay is small. 
In contrast, we find very little contribution from our model for the $\gamma\Lambda$ final state  due to a cancellation of the dominant channels, so that this decay should be dominated by the genuine three-quark component in a more realistic picture of the $\Lambda^*(1520)$ as a hybrid with some three constituent quark component and a substantial meson-baryon cloud.

More precise experimental information and theoretical tools are needed in order to make more quantitative conclusions about the $\Lambda^*(1520)$, but the findings of the present study point in the direction of the $\Lambda^*$ being a composite object of a genuine 3-quark state and a dynamical resonance, with the first component dominating the $\Lambda^*(1520)\to\gamma\Lambda$ decay and the second the $\Lambda^*(1520)\to\gamma\Sigma^0$ decay.
Extra experimental work, measuring other couplings of the $\Lambda^*(1520)$, like the one to $\bar{K}^*$, would also bring relevant information on the nature of the $\Lambda^*(1520)$, as recently shown in Ref. \cite{Hyodo:2006uw}.

\section*{Acknowledgments}
We would like to thank D. Strottman for a thorough reading of the paper.
This work is partly supported by DGICYT contract number
BFM2003-00856 and Generalitat Valenciana, and the E.U. EURIDICE network contract no. 
HPRN-CT-2002-00311. This
research is  part of the EU Integrated Infrastructure Initiative 
Hadron Physics Project
under  contract number RII3-CT-2004-506078.

\end{document}